\documentclass[aps,preprint]{revtex4}
\usepackage{graphicx}
\usepackage{rotating}
\usepackage{dcolumn}
\usepackage{epsfig}

\usepackage{latexsym}
\usepackage{mathrsfs}
\usepackage{amssymb}
\usepackage{amsmath}
\usepackage{amsfonts}

\begin{document}

\title{Onset of Phase Synchronization in Neurons Conneted via Chemical Synapses}

\author{T. Pereira$^1$, M.S. Baptista$^1$, J. Kurths$^1$, M.B. Reyes$^2$}

\address{$^1$Universit{\"a}t Potsdam, Institut f{\"u}r Physik Am Neuen
    Palais 10, D-14469 Potsdam, Deutschland\\ 
$^2$Institute for Nonlinear Science, University of California, San Diego, La Jolla, California 92093-0402
}

\begin{abstract}
  We study the onset of synchronous states in realistic chaotic
  neurons coupled by mutually inhibitory chemical synapses. For the realistic parameters, namely the
  synaptic strength and the intrinsic current, this synapse
  introduces non-coherences in the neuronal dynamics, yet allowing for
  chaotic phase synchronization in a large range of parameters. As we
  increase the synaptic strength, the neurons undergo to a periodic
  state, and no chaotic complete synchronization is found.
\end{abstract}

\maketitle

\section{Introduction}

Many neural networks relay on a balanced configuration of eletrical
and chemical synapses for a normal functioning.  The electrical
synapse is usually associated to processes that require rapid
responses, since the synaptic delay can be negleted. The chemical
one is mediated by means of chemical transmitters and it is usually
associated to processes that do not require rapid responses, since
there is an intrinsic synaptic delay.

Neural networks with electrical synapse, as well as the
analogous linearly coupled oscillators, have recently attracted
much attention mainly because it provides a simple and clear scenario
for the onset of synchronization.  The basic idea behind this is that
for interacting neurons and oscillators with electrical coupling,
increasing the coupling strength leads to synchronous behavior.  This
relation is important since the more synchronous the oscillators are,
the more information between themselves can be exchanged \cite{Murilo-canal}.

On the other hand, this relation to neurons coupled via chemical synapse is
still unclear. In such a synapse, increasing its strength might change
the neuronal dynamics, since the synapse itself is a dynamical system.
This relation becomes even more complex if the chemical synapse is of
the inhibitory type. In that case, while one neuron spikes the synapse
forces the other neuron not to spike.

For eletrical coupling it was found several types of synchronization
in coupled chaotic oscillators.  Complete synchronization
\cite{scomple}, Generalized synchronization \cite{rulkov}.  There is a
type of synchronization which appears for very small coupling strength
the phase synchronization ($PS$) where the coupled chaotic oscillators
have their absolute phase difference bounded but their amplitudes may
be uncorrelated \cite{rose}. It was numerically seen in a variety of
coupled oscillators \cite{reviews}.  These many types of
synchronization were also found in neurons \cite{Elson, Wang-CS} with
eletrical synapses.  In particular, PS was found in two electrically
coupled neurons \cite{Shuai} and in small neural networks
\cite{Wang-SW}.
                                          
The purpouse of this work is to analyze the inhibitory chemical
synapses in coupled chaotic neurons, and its role for synchronization.
We show that there is no complete chaotic synchronization, since as
the coupling increases the neurons undergo to periodic states.  This
offers a great contrast to synapses of the eletrical type in which
complete chaotic synchronization is commomly found.  We also show that
this inhibitory synapse is responsible for introducing phase
synchronous behavior, for a wide range of parameters.  This result is
biologicaly meaninful since the onset of phase synchronization
provides a good enviroment for communication with chaotic systems
since in phase synchronous states one can send information with low
probability of errors\cite{Murilo-canal}.

This paper is organized as follows. In Sec. \ref{neuMod}, we introduce
the realistic Hindmarsh-Rose neuron model and in Sec. \ref{synMod}, we
introduce the model for the inhibitory chemical synapses.  In Sec.
\ref{ps}, we show the likely ocurrance of phase synchronization in two
neurons with chemical synapse and in Sec.  \ref{conclusao}, we give
the conclusions.

\section{Neuron Model}\label{neuMod}

The neurons are described by the Hindmarsh-Rose model which consists
of four coupled differential equations \cite{Pinto}
{\small
\begin{eqnarray}
\dot{x}&=& ay + bx^2 - cx^3 - dz + I \nonumber \\
\dot{y} &=& e - y - fx^2 - gw    \\
\dot{z} &=& \mu(-z + S(x+H)) \nonumber \\
\dot{w} &=& \nu(-kw + r(y+l)) \nonumber 
\end{eqnarray}
}
This model has been shown to be realistic, since it reproduces the
membrane potential of biological neurons \cite{Johnson}, and it is able
to replace a biological neuron in a damaged biological network,
restoring its natural functional activity \cite{Mulle}, it also 
reproduces a series of collective behaviors observed in a living
neural network \cite{Pinto}.   We integrate the Hindmarsh-Rose model using a
Runge-Kutta of order 6 with adaptative step, and set the parameters 
to obtain a spiking/bursting dynamics. The
parameters of the model are: $a=1.0$,$b= 3.0$, $c=1.0$,
$d=0.99$,$e=1.01$,$f=5.0128$,$g=0.0278$,$H=1.605$,
$k=0.9573$,$l=1.619$, $\mu=0.0021$, $\nu=0.0009$, $r=3.000$,
$S=3.966$.

\section{Synapse Model}\label{synMod}

Each synaptic connection between the neurons is modeled by a nonlinear
differential equation that mimics the release of neurotransmitters at
the synaptic cleft and its absorption in the post synaptic cell \cite{Sharp}.
 The current $I_{syn}$ injected in the postsynaptic cell is
determined by the dimensionless, scaled synaptic activation $S(t)$.
{\small
\begin{eqnarray}
I_{syn}(t) &=& g_{syn} S (t) [x(t) - V_{rev}]\\
\tau \frac{d S}{dt} &=& \frac{S_{\infty}(V_{in}(t)) - S(t)}{S_0 - S_{\infty}(V_{in}(t))}, 
\end{eqnarray}
} 
\noindent
where $V_{rev}$ is the synaptic potential, $V_{in}$ is the presynaptic
voltage, $x(t)$ represents the membrane potential of the postsynapict
neuron, and $\tau$ is the timescale governing receptor binding.
$S_{\infty}$ is given by:
{\small
\begin{eqnarray}
S_{\infty}(V) = \left\{
\begin{array}{ccc}
tanh\frac{V - V_{th}}{V_{slope}}, & \mbox{if} & V > V_{th} \\
0 & \mbox{if} & V \leq V_{th} \\
\end{array}
\right.
\end{eqnarray}
}
\noindent
We set the parameters of the synapse equations in order to present an
inhibitory effect. That is done by using the following parameters: $V_{th}=-0.80$,
$V_{slope}= 1.00$,$V_{rev}=-1.58$, and $S_0 \ge 1$.

\section{Phase Synchronization}\label{ps}

The condition for PS can be written as

\begin{equation}
|\phi_1 - q \phi_2| \leq c,
\label{ps_cond_I}
\end{equation}
\noindent
where $\phi_{1,2}$ are the phases calculated from a projection of the
attractor onto appropriate subspaces. The neurons present a
non-coherent dynamics due to the two time scales, i.e.
bursting/spiking behavior.  By non-coherent dynamics we mean that
there is no clear center of rotation in which the trajectory spirals
around and also it is not possible to define a Poincar\'e section for
which the trajectory crosses only once each time the neuron has a
hyperpolarization.

The chemical synapse introduces even more non-coherence. This happens
because when one neuron is in a spiking behavior it inhibits the other
neuron, which might hyperpolarize, but the neuron that has been inhibited
still tries to spike. This competition generates more
non-coherence in the phase space.  As a consequence, it is rather
unclear how one can calculate the phases for such dynamics.  However,
it is possible to overcome this problem by using the conditional
Poincar\'e map \cite{muriloPHD}, which is a map of the attractor,
construct by observing it for specific times at which events occur in
one neuron.  Using such technique, we can detect PS without actually
having to measure the phase.

\subsection{PS-sets}

The conditional Poincar\'e map is a map of the flow. In particular, it
consists in observing the trajectory of the neuron $\mathcal{N}_j$ at
special times $\tau^i_j$, with the index $j=1,2$ indicating the two
neurons.  We define these times of events $\tau^i_j$, by the following
rule:

\begin{itemize}
\item {$\tau^i_1$} represents the time at which the membrane potential
  in the neuron $\mathcal{N}_2$ reaches a threshold for i-th times. 
\item {$\tau^i_2$} represents the time at which the membrane potential
  in the neuron $\mathcal{N}_1$ reaches a threshold for i-th times. 
\end{itemize}
\noindent
Then, we record the trajectory position of the neuron $\mathcal{N}_j$
at these times $\tau^i_j$. As a result, we have a discret set of
points called $\mathcal{D}_j$. If $\mathcal{D}_j$ does not spreads
over the attractor of $\mathcal{N}_j$, but is rather localized, we say
that the set $\mathcal{D}_j$ is a PS-set. It can be shown that PS-set
implies PS \cite{muriloPHD}.  This is so, because the difference
between the time at which the $i$-th event happens in both oscillators
is small, which means that the time difference $|\tau_1^i - \tau_2^i|
< \delta $, with $\delta $ being a small constant.  As a consequence,
the points in the conditional Poincar\'e map are confined.

\begin{figure}[th]
  \centerline{\psfig{file=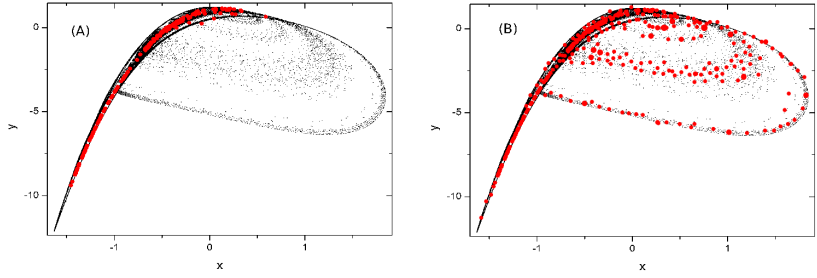,height=4cm}} \vspace*{8pt}
\caption{In black
  it is depicted the attractor and in red the set $\mathcal{D}_j$. In
  (A),  the set $\mathcal{D}_j$ is localized with respect to the attractor, and the map
  $\mathcal{D}_j$ does not fulfill the whole attractor projection, occupying
  just partially this projection. This implies the presence of phase
  synchronization between the two neurons. The parameters are $I =
  3.12$ and $g_{syn}=0.78$. In (B), the set $\mathcal{D}_j$ spreads
  over the attractor and fulfill it, this show that there is no PS,
  the parameters are $I = 3.12$ and $g_{syn}=0.76$ }
\label{ps-set}
\end{figure}

\begin{figure}[th]
\centerline{\hbox{\psfig{file=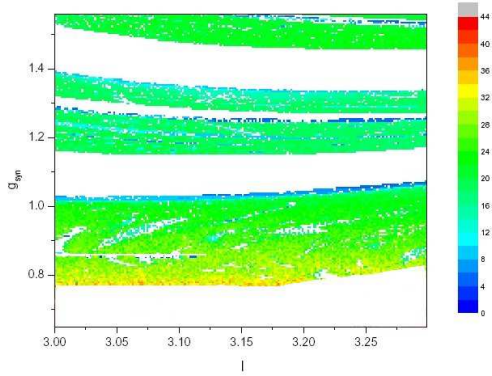,height=5cm}}}
\caption{The Parameter space $I \times g_{sync}$ is
  depicted. The colored regions represent parameters in which the
  neurons present PS-sets, which imply phase synchronization.  White
  regions represent either chaotic or periodic behavior.}
\label{sinc_par}
\end{figure}

\begin{figure}[th]
\centerline{\hbox{\psfig{file=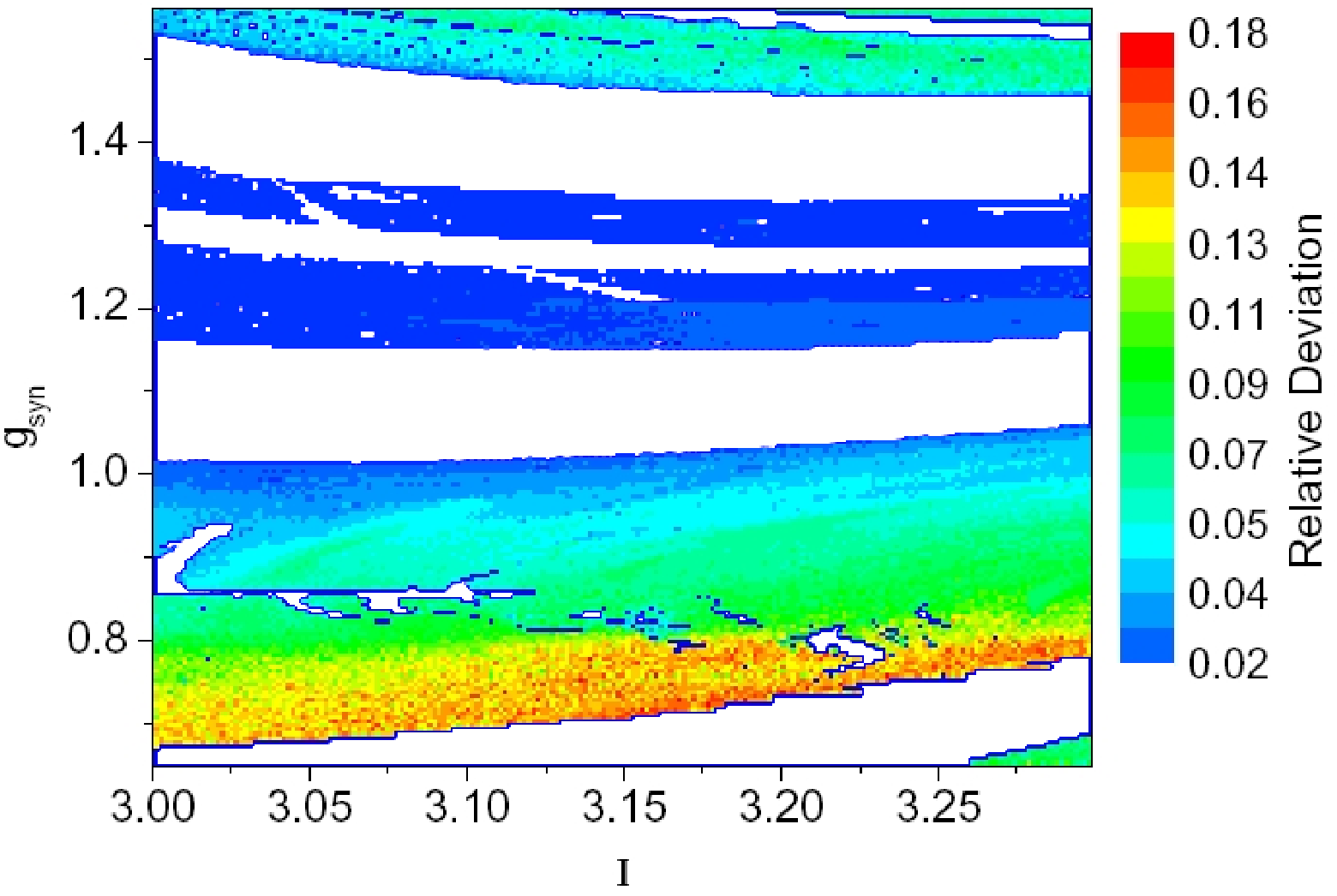,height=5cm}}}
\caption{  The Parameter space $I \times g_{sync}$ is
  depicted. The colored regions represent the relative standard
  deviation of the burst time for one neuron. A positve relative
  standart deviation means that the neuron is still chaotic. The
  greather is the standart deviation the more chaotic the neuron is.
  White regions represent periodic behavior.}
\label{desvio}
\end{figure}

In Fig. \ref{ps-set} we show two types of $\mathcal{D}_j$ set.  In (A), 
the $\mathcal{D}_j$ is a PS-set. One can see that this set is
localized, and so it does not spread over the attractor, then we have
phase synchronization. The parameters are $I = 3.12$ and
$g_{syn}=0.78$. In (B), is a situation where there is no phase
synchronization, for $I = 3.12$ and $g_{syn}=0.76$. The set $\mathcal{D}_j$
spreads over the attractor.
To have a global view of the possible behaviors in this system, in
Fig. \ref{sinc_par} we show the parameter space in the coordinates $I
\times g_{syn}$.  In this parameter space we depicted in color the
parameters for which we have phase synchronization, and in white
parameters for which either chaos with no synchronous behavior or
periodic states is found. The color bar at the right of this figure
indicates the relative area in percentage of the $\mathcal{D}_j$ set
occupation in the attractor projection.  To assure that we have
chaotic phase synchronization we also compute the standard deviation
of the event times. We introduce the quantity $T_j^i = \tau_j^i -
\tau_j^{i-1}$, so its relative standard deviation is given by
$\varsigma = \{ \langle ( T_j^i )^2\rangle - \langle T_j^i \rangle^2
\}/ \langle T_j^i \rangle$, where $\langle \cdot \rangle$ represents
the average. In Fig.  \ref{desvio}, we show $\varsigma$ as a function
of $I$ and $g_{syn}$.
We have two kinds of transitions for phase synchronization. The first
the neurons are chaotic and present a non synchronous behavior then
increasing the synaptical coupling there is a transition to phase
synchronization, fixing the current $I=3.12$ the first transition
happens around $g_{syn}\approx 0.77$.  The second transition the
neurons are in a periodic behavior and when we increase the coupling
strength they undergo to chaos but phase synchronized.
\section{Conclusions}\label{conclusao}
We have shown that phase synchronization is a common behavior in
neurons with inhibitory chemical synapses for the 
Hindmarsh-Rose model.  In addition, it is shown that there is no
complete chaotic synchronization, since as the synapse strength
increases the neurons undergo to periodic states. This places the
inhibitory chemical connection as a good candidate to explain
information transmission processes regulated by means of phase
synchronization.

The neurons present naturally non-coherent dynamics due to the two
time scales provided by the bursting and spiking dynamics.  As they
are coupled by a chemical inhibitory coupling, they undergo to an even
more non-coherent state. However, we could still detect the presence
of phase synchronization using the conditional Poincar\'e map.
\section{Acknowledgments}
We would like to thank R.D. Pinto for useful discussions, and the financial support of
Helmholz Center for Mind and Brain Dynamics (TP and JK), thank Alexander
von Humboldt Foundation (MSB) and  FAPESP (MBR).

\end{document}